\documentclass[sigconf]{acmart}

\usepackage{xspace}

\AtBeginDocument{%
  \providecommand\BibTeX{{%
    \normalfont B\kern-0.5em{\scshape i\kern-0.25em b}\kern-0.8em\TeX}}}

\setcopyright{acmcopyright}
\copyrightyear{2018}
\acmYear{2018}
\acmDOI{10.1145/1122445.1122456}

\acmConference[MSR 2022]{MSR '22: Proceedings of the 19th International Conference on Mining Software Repositories}{May 23–24, 2022}{Pittsburgh, PA, USA}

\newcommand{\dataset}{{\sc Methods2Test}\xspace}
\newcommand{\ie}{\textit{i.e.,}~}
\newcommand{\eg}{\textit{e.g.,}~}

\copyrightyear{2022}
\acmYear{2022}
\setcopyright{acmcopyright}\acmConference[MSR '22]{19th International Conference on Mining Software Repositories}{May 23--24, 2022}{Pittsburgh, PA, USA}
\acmBooktitle{19th International Conference on Mining Software Repositories (MSR '22), May 23--24, 2022, Pittsburgh, PA, USA}
\acmPrice{15.00}
\acmDOI{10.1145/3524842.3528009}
\acmISBN{978-1-4503-9303-4/22/05}

\begin{document}

\title{Methods2Test: A dataset of focal methods mapped to test cases}


\author{Michele Tufano, Shao Kun Deng, Neel Sundaresan, Alexey Svyatkovskiy}
\affiliation{
    \institution{Microsoft}
    \city{Redmond}
    \state{WA}
    \country{USA}
}
\email{Email:{mitufano, shade, neels, alsvyatk}@microsoft.com}

\renewcommand{\shortauthors}{Tufano, et al.}

\begin{abstract}
Unit testing is an essential part of the software development process, which helps to identify issues with source code in early stages of development and prevent regressions. Machine learning has emerged as viable approach to help software developers generate automated unit tests. However, generating reliable unit test cases that are semantically correct and capable of catching software bugs or unintended behavior via machine learning requires large, metadata-rich, datasets.
In this paper we present \dataset: a large, supervised dataset of test cases mapped to corresponding methods under test (\ie focal methods). This dataset contains 780,944 pairs of JUnit tests and focal methods, extracted from a total of 91,385 Java open source projects hosted on GitHub with licenses permitting re-distribution. The main challenge behind the creation of the \dataset~was to establish a reliable mapping between a test case and the relevant focal method. To this aim, we designed a set of heuristics, based on developers' best practices in software testing, which identify the likely focal method for a given test case. To facilitate further analysis, we store a rich set of metadata for each method-test pair in JSON-formatted files. Additionally, we extract textual corpus from the dataset at different context levels, which we provide both in raw and tokenized forms, in order to enable researchers to train and evaluate machine learning models for Automated Test Generation. \dataset~is publicly available at: \url{https://github.com/microsoft/methods2test}
\end{abstract}



\maketitle

\section{Introduction}
Writing test cases is a critical part of software development yet often neglected by developers. While software testing is widely acknowledged as an effective step towards identifying bugs early in the development process, developers tend to prioritize the introduction of new features \cite{daka2014survey} over writing test cases. Automated software testing aims to fill this gap, by providing tools that facilitate the accurate generation, selection, and execution of tests. 

Specifically, in the context of code generation, researchers have proposed several techniques which automate the synthesis of unit test cases, such as EvoSuite~\cite{fraser2011evosuite}, Randoop~\cite{pacheco2007randoop}, and Agitar~\cite{agitar}.

These well-established techniques rely on search-based software engineering, code analysis, program synthesis, and constraint solving to suggest candidate test cases. These approaches often aim at maximizing traditionally accepted metrics such as code coverage (line and branch coverage) and mutation score. The automatically generated test suites can be used for regression testing purposes, providing a relative degree of confidence that the changes performed by developers did not introduce regressions in the system. 

While these techniques undoubtedly provide a valuable contribution for developers, several empirical studies have uncovered weaknesses and limitations of the generated test cases, such as unsatisfactory code quality \cite{palomba2016diffusion, palomba2016automatic, grano2019scented}, poor fault-detection capability \cite{pinto2010multi}, and the inability to adequately meet the software testing needs of industrial developers \cite{almasi2017industrial, shamshiri2015automated}. Moreover, from the developers' perspective, these test cases often appear as machine-generated code, and thus, hard to read, understand, and maintain \cite{daka2015modeling, grano2018empirical}. These limitations may be due to the fact that existing techniques focus mainly on testing metrics such as code coverage and mutation score when generating test cases, rather than learning from real-world, developer-written tests.

Recent years have witnessed the proliferation of machine learning (ML) techniques aiming at solving different tasks within the software engineering (SE) realm. Examples of these tasks include automating developers' activities, such as: bug-fixing \cite{tufano2019empirical,chen2019sequencer}, fault localization \cite{wong2016survey}, writing comments \cite{hu2018deep} and documentation \cite{clement2020pymt5}, code reviews \cite{tufan2021towards}, and merge conflict resolution~\cite{mergebert}. This represents a fundamental shift away from maximizing code metrics and towards learning from real-world examples, extracting patterns from large amount of publicly available data.  

These ML-based techniques require large datasets in order to effectively learn from examples. However, these datasets need to balance the necessity for a large number of examples, with the quality constraints and requirements. As for many other cases, the concept of "\textit{garbage in, garbage out}" applies to these ML-based approaches, which need to be trained on high quality data, since noise could hinder the learning process. Thus, several curated datasets have been proposed for different SE tasks, allowing researchers to train and evaluate their ML-based techniques. Examples of such datasets include Defects4J \cite{just2014defects4j}, Bugs2Fix \cite{tufano2019empirical}, and CodRep \cite{chen2019sequencer} for automated bug-fixing,  NASA Metrics Data Program (MDP) \cite{shepperd2013data}, PROMISE \cite{promise}, and the unified dataset \cite{ferenc2018public} for defect prediction, Landfill \cite{palomba2015landfill} for code smell detection, and many others. 

In the field of automated test generation, however, there is a noticeable absence of a large, curated dataset of real world test cases. Additionally, in order to effectively train an ML-based approach to generate test cases, the dataset cannot be a simple collection of tests, rather it needs to expose traceability links between tests and their corresponding methods under tests (\ie focal methods). 

To fill this gap, we propose \dataset: a large dataset of test cases mapped to their corresponding focal methods. This dataset contains 780k pairs of JUnit tests and focal methods, extracted from a total of 91k Java open source projects hosted on GitHub. In order to establish a reliable mapping between a test case and the relevant method under test, we designed a set of heuristics, based on developers' best practices in software testing, such as class and method naming conventions.

To facilitate and enable further analysis even beyond the scope we initially intended, we store a rich set of metadata for each method-test pair in JSON-formatted files, including information about project, file, class, and methods related to the test case. Additionally, we extract parallel textual corpus from the dataset at different context levels, which we provide both in raw and tokenized forms, in order to allow researchers to train and evaluate ML models for Automated Test Generation. \dataset~data and scripts are publicly available at: \url{https://github.com/microsoft/methods2test}

\section{Dataset Generation}
This section provides the details on how we created \dataset. We begin by describing the steps for mining test cases, and mapping them to corresponding focal methods (Sec. \ref{sec:mapping}). Then we extract the different levels of focal contexts surrounding the focal method, which provides a useful representation for a test case generation model (Sec. \ref{sec:focal_context}). Finally, we illustrate how we organize and store the dataset (Sec. \ref{sec:storage}).

\subsection{Test Case Extraction \& Mapping}
\label{sec:mapping}
The goal of this stage is to mine test cases and corresponding focal methods (\ie the method tested by the test case) from a set of Java projects. We select a 91K sample of all the public GitHub Java repositories declaring an open source license, which have been updated within the last five years, and are not forks.

First, we parse each project to obtain classes and methods with their associated metadata. Next, we identify each test class and its corresponding focal class. Finally, for each test case within a test class, we map it to the related focal method obtaining a set of mapped test cases.

\subsubsection*{Parsing}
We parse each project under analysis with the \texttt{tree-sitter} parser\cite{treesitter}. During the parsing, we automatically collect metadata associated with the classes and methods identified within the project. Specifically, we extract information such as method and class names, signatures, bodies, annotations, and variables. The parsed code is used to identify test cases and corresponding focal methods, as well as augmenting the focal methods with focal context.

\subsubsection*{Finding Test Classes}
In this stage, we identify all the test classes, which are classes that contain a test case. To do so, we mark a class as a test class if it contains at least one method with the \texttt{@Test} annotation. This annotation informs JUnit that the method to which it is attached can be run as a test case.

\subsubsection*{Finding Focal Classes}
For each test class we aim to identify the focal class which represents the class under test. To this aim, we employ the following two heuristics, in a sequence:

\begin{itemize}
\item \textit{Path Matching}: best practices for JUnit testing suggest placing code and corresponding test cases in a mirrored folder structure. Specifically, given the class \texttt{src/main/java/Foo.java} the corresponding JUnit test cases should be placed in the class \texttt{src/test/java/FooTest.java}. Our first heuristic tries to identify the folder where the focal class is defined, by following the path of the test class but starting with the \texttt{src/main} folder (\ie production code).

\item \textit{Name Matching}: the name of a test class is usually composed of the name of the focal class, along with a "Test" prefix or suffix. For example, the test case for the class \texttt{Foo.java} would probably be named \texttt{FooTest.java}. Thus, following the path matching heuristic, we perform name matching to identify the focal class by matching the name of the test case without the (optional) "Test" prefix/suffix.
\end{itemize}


  
  

  
    
        



\subsubsection*{Finding Focal Method}
For each test case (\ie method within a test class with the \texttt{@Test} annotation) we attempt to identify the corresponding focal method within the focal class. To this aim, we employ the following heuristics:

\begin{itemize}
\item \textit{Name Matching}: following the best practices for naming classes, test case names are often similar to the corresponding focal methods. Thus, the first heuristic attempts to match the test cases with a focal method having a name that matches, after removing possible \texttt{Test} prefix/suffix. 

\item \textit{Unique Method Call}: if the previous heuristic did not identify any focal method, we compute the intersection between (i) the list of method invocations within the test case and (ii) the list of methods defined within the focal class. If the intersection yields a unique method, then we select the method as the focal method. The rationale behind this approach is as follows: since we have already matched the test class with the focal class (with very high confidence heuristics), if the test case invokes a single method within that focal class, it is very likely testing that single method.

\end{itemize}

\subsubsection*{Mapped Test Cases}
The result of the data collection phase is a set of mapped test cases, where each test case is mapped to the corresponding focal method. It is important to note that we discard test cases for which we were not able to identify the focal method using our heuristics. We designed these heuristics to be based on testing best practices, and obtain a correct mapping with very high confidence. This will likely exclude test cases that have been automatically generated, which may not follow testing best practices.

We collect an initial set of 887,646 mapped test case pairs. From this set, we exclude duplicates, remaining with a total of 780,944 unique mapped test case pairs. Next, we split the dataset into training ($\sim$80\% - 624,022 pairs), validation ($\sim$10\% - 78,534 pairs), and test ($\sim$10\% - 78,388 pairs) sets. We performed this split by carefully taking into account possible data leakage. Specifically, during the split we enforce the constraint that any two data points belonging to the same repository cannot be placed in two different sets (\eg one in training and the other in test). That is, all the data points belonging to the same repository will be placed in the same set. 

Table \ref{tab:methods2test} reports the details of the dataset split, with number of repositories and mapped test cases.

\begin{table}[t]
\centering
	\vspace{-0.0cm}
	\caption{\dataset~Dataset}
	\label{tab:methods2test}
	\resizebox{0.8\linewidth}{!}{
\begin{tabular}{lrr}
\toprule
Set & Repositories & Mapped Test Cases\\
\midrule
Training & 72,188 & 624,022 \\
Validation & 9,104 & 78,534 \\
Test & 10,093 & 78,388 \\
\midrule
Total & 91,385 & 780,944 \\
\bottomrule
\end{tabular}}
\vspace{-0.4cm}
\end{table}

\subsection{Focal Context}
\label{sec:focal_context}
The goal of this phase is to construct an input which contains the necessary information that an ML-based model could leverage to automatically generate correct and useful test cases. Intuitively, the focal method (\ie the method under test) represents the core information to feed to a model. However, additional contextual information can provide important clues for the model to better understand the focal method nature and its context, improving the likelihood of generating test cases that compile and properly test the focal method.

We build different versions of the code input representation -- with diverse degree of focal context -- with the aim of providing to researchers and practitioners, a variety of corpora, of different lengths, that can be used to train their models. We begin with the core information (\ie focal method) and iteratively add contextual information such as class name, constructors, other method signatures, and fields.

\begin{figure}[t!]
    \centering
    \vspace{-0.4cm}
    \caption{Focal Context}
    \vspace{0.1cm}
    \includegraphics[width=0.47\textwidth]{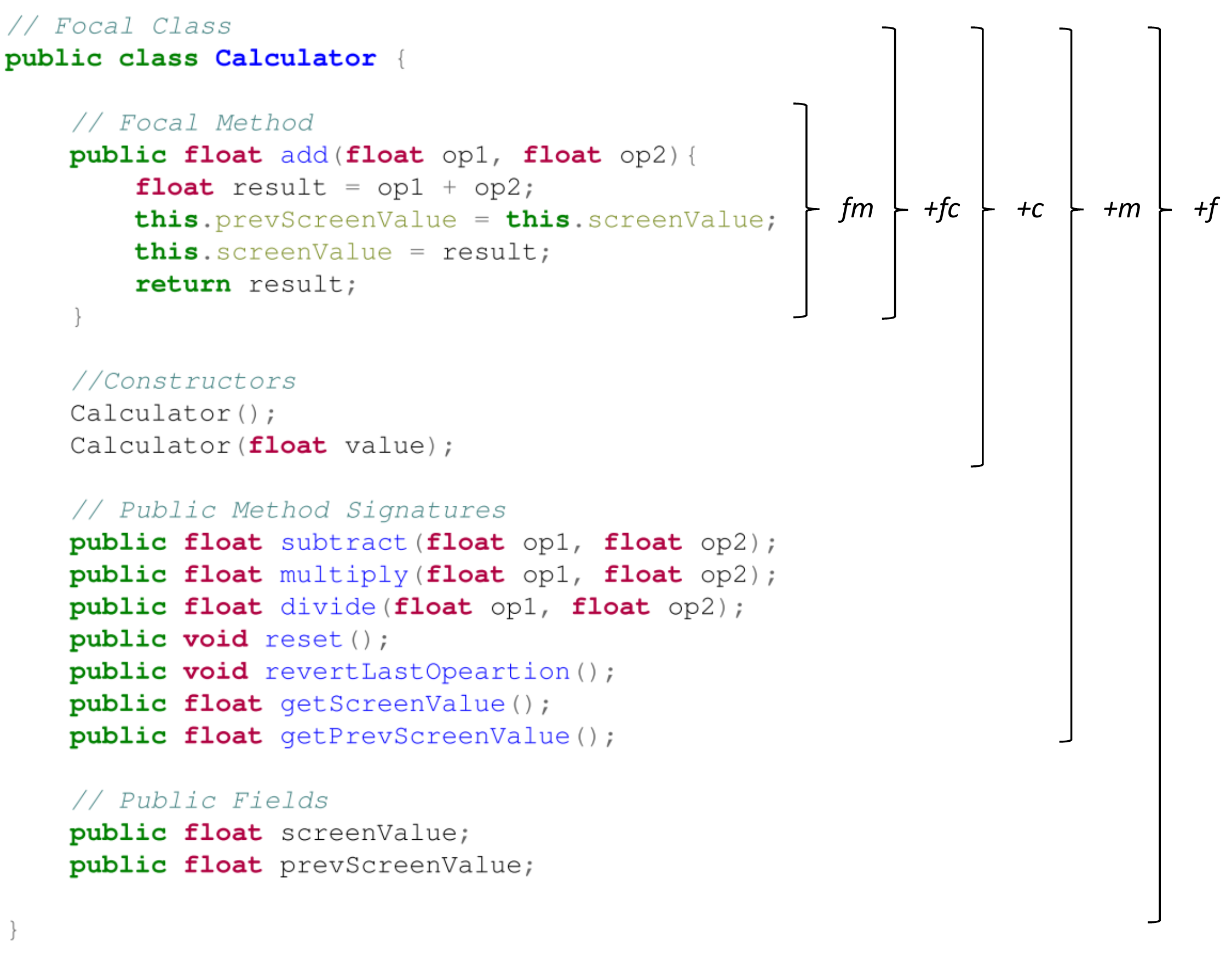}
    \vspace{-0.5cm}
    \label{fig:focal-context}
\end{figure}

Figure \ref{fig:focal-context} provides an overview of the different levels of context we generate for the focal method \texttt{add} in the \texttt{Calculator} class. The left side corresponds to the textual representation, while the right side delineates the context which is indicated with a focal context ID, which we describe in the following:

\begin{itemize}
    \item \textit{fm}: this representation incorporates exclusively the source code of the focal method. Intuitively, this contains the most important information for generating accurate test cases for the given method.
    
    \item \textit{fm+fc}: this representations adds the focal class name, which can provide meaningful semantic information to the model.
    
    \item \textit{fm+fc+c}: this representation adds the signatures of the constructor methods of the focal class. The idea behind this augmentation is that the test case may require instantiating an object of the focal class in order to properly test the focal method.
    
    \item \textit{fm+fc+c+m}: this representation adds the signatures of the other public methods in the focal class. The rationale which motivated this inclusion is that the test case may need to invoke other auxiliary methods within the class (\eg getters, setters) to set up or tear down the testing environment.
    
    \item \textit{fm+fc+c+m+f}: this representation adds the public fields of the focal class. The motivation is that test cases may need to inspect the status of the public fields to properly test a focal method.
\end{itemize}

 While constructing these representations we face two opposing goals: (i) include as many tokens as possible, given their powerful expressiveness discussed above (ii) keep a concise representation that fits into GPU memory, allowing anyone to train their model.

Intuitively, having a representation that includes many tokens from the focal context allows a model to \textit{attend} to different parts of the input and leverage these information to generate a correct and meaningful test case. On the other hand, irrelevant tokens could represent noise for the learning process, which could lead to worse performances, as well as wasting GPU memory that could be use for more informative tokens.

It is important to highlight that in our representation, the order of inclusion of a particular focal context, for example the constructors' signatures (\textit{fm+fc+c}) before other methods' signatures (\textit{fm+fc+c+m}), is important, since the textual representation could be truncated if it exceeds 1024 tokens (\ie maximum sequence length in most Transformer models). 

This order of inclusion has been defined by the authors based on their understanding and intuition of the meaningful clues for test case generation within the focal class. We empirically evaluated these design decision in our previous work \cite{tufano2021unit}.

\subsection{Organization \& Storage}
\label{sec:storage}
\dataset~is publicly available on GitHub\cite{methods2test}. The repository makes use of the Git large file storage (LFS) service. Git LFS works by replacing large files in the repository with tiny pointer files.

We organize \dataset~in three main folders: dataset, corpus, scripts.
The \texttt{dataset} folder contains the test cases mapped to their corresponding focal methods, along with a rich set of metadata. The dataset is stored as JSON files for each individual method-test pair. Each JSON file contains metadata at repository-, class-, and method-level for both the test case and the focal method. The repository data include:
\begin{itemize}
    \item id: unique identifier of the repository in the dataset
    \item  url: repository URL
    \item  language: programming languages of the repository
    \item  is\_fork: boolean, whether repository is a fork
    \item  fork\_count: number of forks
    \item  stargazer\_count: cumulative number of stars on GitHub
\end{itemize}

The class-level information contains the following fields, both for the focal class and the test class:
\begin{itemize}
    \item identifier: class name
    \item superclass: superclass definition
    \item interfaces: interface definition
    \item fields: list of class fields
    \item methods: list of class methods
    \item file: relative path
\end{itemize}

The method-level information contains the following fields, both for the focal method and test case:
\begin{itemize}
    \item identifier: method name 
    \item parameters: parameter list of the method
    \item body: source code of the method
    \item signature: method signature
    \item testcase: boolean, whether the method is a test case
    \item constructor: boolean, whether the method is a constructor
    \item invocations: list of all methods invoked in the file
\end{itemize}

The \texttt{corpus} folder contains the parallel corpus of focal methods and test cases, as json, raw, tokenized, and preprocessed, suitable for training and evaluation of the model. The corpus is organized in different levels of focal context (described in Sec. \ref{sec:focal_context}), incorporating information from the focal method and class within the input sentence, which can inform the model when generating test cases. We use a  Byte-Level BPE tokenizer to tokenize the corpus, and fairseq to preprocess (binarize) the tokenized corpus.

The \texttt{scripts} folder contains the scripts to mine test cases and map them to the corresponding focal methods. The GitHub repository contains detailed instructions and additional information. 

We created a persistent identifier (DOI), archiving our GitHub repository on Zenodo: \url{https://zenodo.org/badge/latestdoi/451656023}


\section{Applications}
The main goal of \dataset~is to enable researcher to train and evaluate ML-based models aiming at automatically generate unit test cases. We initially build this dataset in conjunction with our previous work \cite{tufano2021unit}, where we proposed AthenaTest, an Encoder-Decoder Transformer model trained to translate a focal method (with focal context) into the corresponding test case. Since AthenaTest is trained on real-world test cases, it is able to generate realistic and readable tests, which resemble those written by developers. 

\dataset~provides the corpus at different levels of focal context, which allows not only to replicate AthenaTest, but also the training of models of different sizes (\eg smaller models on smaller representation). Additionally, the availability of rich metadata in JSON format supports the augmentation of the focal context with novel information, as well as reorganizing the representation in custom ways, differently from what we proposed.

Applications of this dataset may extend beyond what we initially intended. Several empirical studies could be performed on this data, leveraging the traceability between test cases and functional code.

\section{Limitations and Extensions}
A major limitation and threat to validity of \dataset~could arise from noisy data within the dataset. Specifically, incorrect mapping between tests and focal methods could introduce erroneous data points which can hinder the learning process when training models. To mitigate this threat, we rely on naming heuristics based on testing best practices, aiming at collecting only test-method pairs with high confidence. We validate our heuristics by inspecting a statistically significant sample (confidence level of 95\% within 10\% margin of error) of 97 samples from the training set. Two authors independently evaluated the sample, then met to discuss the disagreements. We found that 90.72\% of the samples have a correct link between the test case and the corresponding focal method.

This dataset currently only contains JUnit test cases, limiting the analysis on Java projects which relies on JUnit framework. We are currently working on extending this dataset including projects in different languages (\eg python, c\#) and supporting multiple testing frameworks.

Finally, we would like to extend this dataset with runtime and coverage information attached to each test case. This would require to build each software project, execute test cases, collect runtime information (\ie executed, passed, failed), and compute line/branch coverage on the focal method as well as auxiliary methods.

\section{Related Work}
Our work is related to a collection of dataset and studies focusing on software testing. The most related dataset we found is TestRoutes \cite{kicsi2020testroutes}, where the authors manually curated a test-to-code traceability dataset containing the traceability information on 220 test cases. While we share a similar goal, \dataset~was intended to be a large-scale repository of test cases to enable training of ML-based techniques for automated test generation. Thus, we designed heuristics to perform such mapping, rather than relying on manual analysis. 

Defects4J \cite{just2014defects4j} offers a collection of reproducible bugs, as well as the triggering tests for each of these bugs. The link between the bug and the triggering test could potentially be used to recover the focal method. Defects4J represents an invaluable infrastructure for APR techniques, however, in terms of test cases, they only belong to 17 Java projects, while we mine tests from thousands of open source projects hosted on GitHub.

\section{Conclusions}
We presented \dataset: a large dataset of 780k JUnit test cases mapped to their corresponding focal methods (\ie method under test), extracted from 91k Java open source projects hosted on GitHub. We described the heuristics, based on testing best practices, to extract the test cases and identify the corresponding focal methods. We collect a rich set of metadata for each method-test pair, and store them in JSON-formatted files. Additionally, we extract textual corpus from the dataset at different context levels, which we provide both in raw and tokenized forms, in order to enable researchers to train and evaluate ML models for Automated Test Generation. \dataset~is publicly available at: \url{https://github.com/microsoft/methods2test}

\balance

\bibliographystyle{ACM-Reference-Format}
\bibliography{ms}

\end{document}